\documentclass{article}
\usepackage[utf8]{inputenc}
\usepackage{geometry}
\usepackage{amsmath,amsfonts}
\usepackage{algorithmic}
\usepackage{algorithm}
\usepackage{array}
\usepackage{textcomp}
\usepackage{stfloats}
\usepackage{url}
\usepackage{cite}
\usepackage{verbatim}
\usepackage[pdftex]{graphicx}
\usepackage{color}
\usepackage{transparent}
\usepackage{import}
\usepackage{subfigure}
\usepackage{multirow}
\usepackage{cleveref}
\DeclareGraphicsExtensions{.eps}
\usepackage{epstopdf}
\usepackage{epsfig}
\usepackage{gensymb}
\usepackage{float} 
\usepackage{arydshln} 
 \usepackage{graphicx}
 \usepackage{titling}

 \title{Study of 5G base station antenna array performance for self-interference reduction  
}
\author{Bing Xue}
\date{December 2022}
 
 \usepackage{fancyhdr}
\fancypagestyle{plain}{
    \fancyhf{} 
    \fancyfoot[R]{\includegraphics[width=2cm]{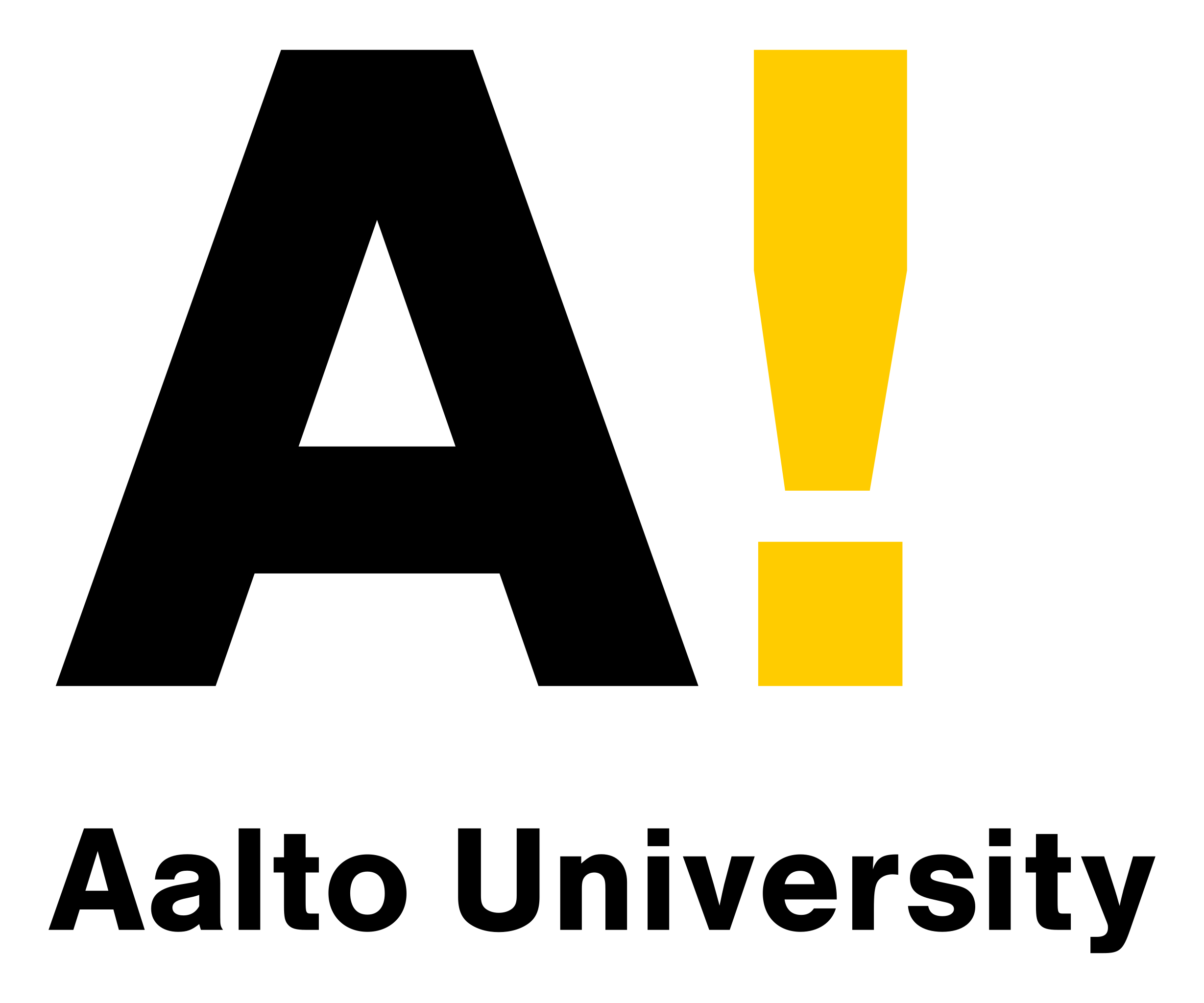}}
    \fancyfoot[L]{\thedate}
    \fancyhead[L]{Description of Algorithm}
    \fancyhead[R]{\theauthor}
}

\makeatletter
\def\@maketitle{%
  \newpage
  \null
  \vskip 1em%
  \begin{center}%
  \let \footnote \thanks
    {\LARGE \@title \par}%
    \vskip 1em%
  \end{center}%
  \par
  \vskip 1em}
\makeatother

\usepackage{lipsum}  
\usepackage{cmbright}

\begin{document}

\maketitle

\noindent\begin{tabular}{@{}ll}
    First Author & \theauthor\\
    Second Author &  Haneda Katsuyuki\\
     Third Author &  Clemens Icheln
\end{tabular}

\section*{Only Line-of-sight Signals' Channel Model}
We can obtain the complex radiation patterns ($E_{\rm 0}(\theta,\phi)$) of all antenna elements on an antenna array over a reference sphere with the radius $d_0^{s}$, where $\theta$ and $\phi$ are the coordinates in the spherical coordinate system. We divide the patterns of all antenna elements into two parts, i.e., $E_0^\mathrm{d}$ and $E_0^\mathrm{u}$. In this way, we can easily estimate the ideal downlink and uplink channels using ideal receiving antennas and transmitting antennas.

First, the receiving signal at the downlink user from the $i$th antenna at any location $(\theta_0,\phi_0,d_i^\mathrm{d})$ can be estimated by geometric optics. 
\begin{equation}
\label{freeprogation1}
{
E_{\mathrm{r},i}^\mathrm{d}(\theta_0,\phi_0,d_i^\mathrm{d}) = E_{0,i}^\mathrm{d}(\theta_0,\phi_0,d_0^\mathrm{s})\frac{d_0^{s}}
{d_i^\mathrm{d}} \exp{\left(-\mathrm{j}2\pi\frac{d_i^\mathrm{d}-d_0^{s}}{\lambda_0}\right)},
}
\end{equation}
where $\mathrm{j} = \sqrt{-1}$ and $\lambda_0$ is the wavelength of the channel central frequency; $d_i^\mathrm{d}$ is the distance between the base-station transmitting antenna and the downlink user; $E_{0,i}^\mathrm{d}$ is also the far field at the location $(\theta_0,\phi_0,d_0^\mathrm{s})$ obtained by full-wave simulations in commercial software, i.e., \textit{CST studio suite}, where $d_0^{s}=1 \mathrm{m}$.

Then, by considering that the two-antenna transmission system is reciprocal, the receiving signal at the $i$th base-station antenna from the uplink user at any direction $(\theta_0,\phi_0,d_i^\mathrm{u})$ can be estimated as
\begin{equation}
\label{freeprogation2}
{
E_{\mathrm{r},i}^\mathrm{u}(\theta_0,\phi_0,d_i^\mathrm{u}) = E_{0,i}^\mathrm{u}(\theta_0,\phi_0,d_0^\mathrm{s})\frac{d_0^{s}}
{d_i^\mathrm{u}} \exp{\left(-\mathrm{j}2\pi\frac{d_i^\mathrm{u}-d_0^{s}}{\lambda_0}\right)},
}
\end{equation}
where $d_i^\mathrm{u}$ is the distance between the base-station receiving antenna and the uplink user; $E_{0,i}^\mathrm{u}$ is also the far field at the location $(\theta_0,\phi_0,d_0^\mathrm{s})$ obtained by full-wave simulations, too.

If there are only line-of-sight signals, assume that the received power at the $i$th base-station antenna from the feeding ports is $P_{0,i}$. By Friis transmission formula, we can get that
\begin{equation}
\label{Friis}
{
P_{\mathrm{r},i}(\theta_0,\phi_0) =P_{0,i}\frac{\lambda_0^2 G_0(\theta_0,\phi_0)}
{(4\pi d_i)^2},
}
\end{equation}
where $G_0$ can be represented by 
\begin{equation}
\label{Gain}
{
G_{0,i}(\theta_0,\phi_0) = \left\|E_{0,i}(\theta_0,\phi_0)\right\|^2 \frac{2\pi{(d_0^{s})}^2}{P_{0,i} \eta_0 },
}
\end{equation}
in which $\eta_0$ is wave impedance in free space. In addition, $P_{\mathrm{r},i}(\theta_0,\phi_0)$ can be written as

\begin{equation}
\label{Power}
{
P_{\mathrm{r},i}(\theta_0,\phi_0) =\left\|E_{r,i}(\theta_0,\phi_0)\right\|^2\frac{\lambda_0^2}{8\pi\eta_0},
}
\end{equation}

Therefore, we can estimate the downlink and uplink channels by combining (\ref{freeprogation1}) or (\ref{freeprogation2}) and (\ref{Friis}). 
If we define 
\begin{equation}
\label{Hdefinition}
{
h_i = \frac{\left\|v_\mathrm{r}\right\|}
{\left\|v_0\right\|} \exp{\left(\mathrm{j}(\Phi_{\mathrm{r}}-\Phi_\gamma)\right)},
}
\end{equation}
where $v_\mathrm{r}$ is the receiving voltage at the user's ideal antenna,  and $v_0$ is the transmitting voltage at an antenna in a base station with a reference phase $\Phi_\gamma=0$, the estimated channels can be written as
\begin{equation}
\label{Channelup}
{
h_{\mathrm{up},i}(\theta_0,\phi_0,d_i^\mathrm{u}) = \sqrt{G_{0,i}(\theta_0,\phi_0)}\frac{\lambda_0}
{4\pi d_i^\mathrm{u}} \exp{\left(\mathrm{j}(\Phi_{0,i}^u(\theta_0,\phi_0,d_0^\mathrm{s})+\Phi_{\delta_\mathrm{u}})-\mathrm{j}2\pi\frac{d_i^\mathrm{u}-d_0^{s}}{\lambda_0}\right)},
}
\end{equation}
and
\begin{equation}
\label{Channeldown}
{
h_{\mathrm{down},i}(\theta_0,\phi_0,d_i^\mathrm{d}) = \sqrt{G_{0,i}(\theta_0,\phi_0)}\frac{\lambda_0}
{4\pi d_i^\mathrm{d}} \exp{\left(\mathrm{j}(\Phi_{0,i}^d(\theta_0,\phi_0,d_0^\mathrm{s})+\Phi_{\delta_\mathrm{d}})-\mathrm{j}2\pi\frac{d_i^\mathrm{d}-d_0^{s}}{\lambda_0}\right)},
}
\end{equation}
where $h_{\mathrm{up},i}(\theta_0,\phi_0,d_i^\mathrm{u}) = C_\mathrm{up} E_{\mathrm{up},i}^\mathrm{G}(\theta_0,\phi_0)\frac{\exp{\left(-\mathrm{j}2\pi\frac{d_i^\mathrm{u}}{\lambda_0}\right)}}{d_i^\mathrm{u}}$ is the uplink channel for the $i$th base station antenna when the user is at $(\theta_0,\phi_0,d_i^\mathrm{u})$, and 
$h_{\mathrm{down},i}(\theta_0,\phi_0,d_i^\mathrm{d})=C_\mathrm{down} E_{\mathrm{down },i}^\mathrm{G}(\theta_0,\phi_0)\frac{\exp{\left(-\mathrm{j}2\pi\frac{d_i^\mathrm{d}}{\lambda_0}\right)}}{d_i^\mathrm{d}}$ is the downlink channel for the $i$th base station antenna when the user is at $(\theta_0,\phi_0,d_i^\mathrm{d})$;
$E_{0,i}^\mathrm{u}(\theta_0,\phi_0,d_0^\mathrm{s}) = \left\|  E_{0,i}^\mathrm{u}(\theta_0,\phi_0,d_0^\mathrm{s}) \right\| \exp{\left(\mathrm{j}\Phi_{0,i}^\mathrm{u}(\theta_0,\phi_0,d_0^\mathrm{s})\right)}$ and 
$E_{0,i}^\mathrm{d}(\theta_0,\phi_0,d_0^\mathrm{s}) = \left\|  E_{0,i}^\mathrm{d}(\theta_0,\phi_0,d_0^\mathrm{s}) \right\| \exp{\left(\mathrm{j}\Phi_{0,i}^\mathrm{d}(\theta_0,\phi_0,d_0^\mathrm{s})\right)}$;
The complex gain patterns at the coordinate origin are $E_{\mathrm{up},i}^\mathrm{G}(\theta_0,\phi_0) = \sqrt{G_{0,i}(\theta_0,\phi_0)}\exp{\left(\mathrm{j}\Phi_{0,i}^\mathrm{u} -\mathrm{j}2\pi\frac{d_0^\mathrm{s}}{\lambda_0}\right)}$ and $E_{\mathrm{down},i}^\mathrm{G}(\theta_0,\phi_0) = \sqrt{G_{0,i}(\theta_0,\phi_0)}\exp{\left(\mathrm{j}\Phi_{0,i}^\mathrm{d} -\mathrm{j}2\pi\frac{d_0^\mathrm{s}}{\lambda_0}\right)}$;
$\Phi_{\delta_\mathrm{d}}$ and $\Phi_{\delta_\mathrm{u}}$ are the shifted phases due to the transformation from a free-space wave to a guided wave on feeding ports of ideal user antennas. 

In a matrix form, the uplink channel is $\boldsymbol{H}_{\mathrm{up}}= \left[ \boldsymbol{H}_{\mathrm{up,1}}^\mathsf{T},\boldsymbol{H}_{\mathrm{up,2}}^\mathsf{T},...,\boldsymbol{H}_{\mathrm{up},M_\mathrm{up}}^\mathsf{T} \right]^\mathsf{T}$, while the downlink channel is $\boldsymbol{H}_{\mathrm{down}} = \left[ \boldsymbol{H}_{\mathrm{down,1}},\boldsymbol{H}_{\mathrm{down,2}},...,\boldsymbol{H}_{\mathrm{down},M_\mathrm{down}} \right]$, in which $M_\mathrm{up}+M_\mathrm{down}\leq\mathrm{M}$, and $\rm M$ is the number of antenna elements in a base-station array.; $\boldsymbol{H}_{\mathrm{up},i} = \left[ h_{\mathrm{up},i1}(\theta_{i1},\phi_{i1},d_{i1}^\mathrm{u}) ,h_{\mathrm{up},{i2}}(\theta_{i2},\phi_{i2},d_{i2}^\mathrm{u}) ,..., h_{\mathrm{up},{ij}}(\theta_{ij},\phi_{ij},d_{ij}^\mathrm{u}) \right]$ and $\boldsymbol{H}_{\mathrm{down},i} = \left[ h_{\mathrm{down},i1}(\theta_{i1},\phi_{i1},d_{i1}^\mathrm{d}) ,h_{\mathrm{down},{i2}}(\theta_{i2},\phi_{i2},d_{i2}^\mathrm{d}),...,h_{\mathrm{down},{ij}}(\theta_{ij},\phi_{ij},d_{ij}^\mathrm{d}) \right]^\mathsf{T}$ where  $i \leq M_\mathrm{up}\& M_\mathrm{down}$ and $j \leq K_\mathrm{up}\& K_\mathrm{down}$ are a positive integer; $\boldsymbol{H}_{\mathrm{up}}\in\mathbb{C}^{M_\mathrm{up} \times K_\mathrm{up}}$ and $\boldsymbol{H}_{\mathrm{down}}\in\mathbb{C}^{K_\mathrm{down} \times M_\mathrm{down}}$.
\newline

For a planar uniform antenna array (uplink), assume the $M_{\mathrm{up,x}}$ elements along the $x$ axis with spacing $a$ and $M_{\mathrm{up,y}}$ elements along the $y$ axis with spacing $b$, so that there are $M_{\mathrm{up}} =M_{\mathrm{up,x}} \times M_{\mathrm{up,y}}$ elements in total. $k \leq M_{\mathrm{up}}$ is a positive integer. Then, $i_k=k\mod M$ is the $x$ index of element $k$ and $j_k = 1 + \mathrm{int}((k-1)/M_{\mathrm{up,x}})$ is the $y$ index of element $k$. $\kappa$ is the wave number in free space. The practical communication environments satisfy $\kappa d \gg 1 $, in which $d$ is the distance between the base-station antenna array and the user so that every antenna receives the same signal nearly in the same direction. Assume the 
$h_{\mathrm{up},k}(\theta_{\gamma},\phi_{\gamma},d_{\gamma}^\mathrm{u}) = C_\mathrm{up} E_{\mathrm{up},k}^\mathrm{G}(\theta_\gamma,\phi_\gamma)\frac{\exp{\left(-\mathrm{j}\kappa d_{\gamma}^\mathrm{u}\right)}}{d_{\gamma}^\mathrm{u}}$ for the $k$th antenna element and the $\gamma$th user. 

\section*{Rayleigh Channel Model}
Similar to LOS signal channel model, we continue using $\boldsymbol{H}_{\mathrm{up}}= \left[ \boldsymbol{H}_{\mathrm{up,1}}^\mathsf{T},\boldsymbol{H}_{\mathrm{up,2}}^\mathsf{T},...,\boldsymbol{H}_{\mathrm{up},M_\mathrm{up}}^\mathsf{T} \right]^\mathsf{T}$. We assume that the signals from users come equally from all angles over a sphere. The signals' strength meets the Rayleigh fading. In other words, these signals ($p(\theta,\phi) = \frac{1}{\sqrt{2}}(x + \mathrm{j}y)$) are independent and identically distributed (i.i.d.) zero-mean complex Gaussian variables with unit variance. Therefore, we modify $h_{\mathrm{up},k}$ to obtain \cite{Quist2009}
\begin{equation}
\label{Rayleighcomponent}
{
\hat{h}_{\mathrm{up},k}(d_{\gamma}^\mathrm{u}) = C_\mathrm{up}\frac{\exp{\left(-\mathrm{j}\kappa d_{\gamma}^\mathrm{u}\right)}}{d_{\gamma}^\mathrm{u}} \int {E_{\mathrm{up},k}^\mathrm{G}(\theta,\phi)}p(\theta,\phi) \sin{\theta} \mathrm{d}\theta \mathrm{d}\phi
}
\end{equation}
in which the slow fading and the shadowing in large-scale fading are not included.

\begin{figure}[htbp] 
    \centering
	\includegraphics[width=0.5\linewidth]{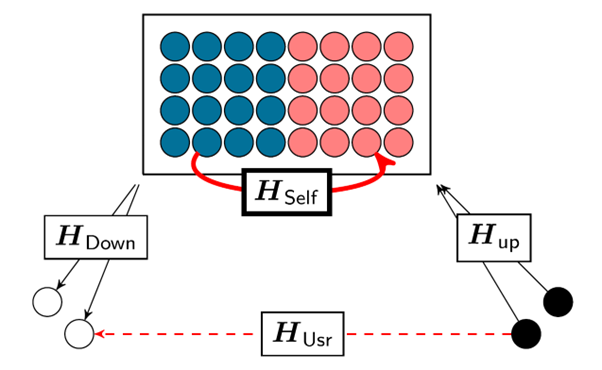}
	\caption{Multi-user full-duplex system}
	\label{onehandgain}
\end{figure}

\section*{Self-interference Reduction (Consider one precoder)}
The signal received at the base station is 
\begin{equation}
\label{receiving_up2}
{
\boldsymbol{y}_{\mathrm{up}} = \boldsymbol{H}_{\mathrm{up}}\boldsymbol{x}_{\mathrm{up}}+\boldsymbol{H}_{\mathrm{self}}\boldsymbol{x}_{\mathrm{down}} + \boldsymbol{n}_\mathrm{up},
}
\end{equation}
where $\boldsymbol{H}_{\mathrm{self}}\in\mathbb{C}^{M_\mathrm{up}\times M_\mathrm{down}}$, which is an mutual-coupling matrix, and $\boldsymbol{n}_\mathrm{up} \in \mathbb{C}^{M_\mathrm{up}}$ is the noise at the base-station receiving antennas. We can obtain all antenna elements' reflection ($S_{ii}$) and mutual-coupling coefficients ($S_{ij}$); where $i\&j\leq \rm M$ is a positive integer and $i\neq j$. The signals received by the downlink users are 
\begin{equation}
\label{receiving_down2}
{
\boldsymbol{y}_{\mathrm{down}} = \boldsymbol{H}_{\mathrm{down}}\boldsymbol{x}_{\mathrm{down}}+\boldsymbol{H}_{\mathrm{user}}\boldsymbol{x}_{\mathrm{up}} +\boldsymbol{n}_\mathrm{down},
}
\end{equation}
where $\boldsymbol{H}_{\mathrm{user}} = 0$ in our simplifying assumption; $\boldsymbol{n}_\mathrm{down} \in \mathbb{C}^{K_\mathrm{down}}$ is the noise at $K_\mathrm{down}$ downlink users. Then, we define the precoders, $\boldsymbol{P}_\mathrm{r}$ and $\boldsymbol{P}_\mathrm{t}$ \cite{Riihonen2011}. 
\begin{equation}
\label{receiving_upNew}
{
\hat{\boldsymbol{y}}_{\mathrm{up}} = \boldsymbol{P}_\mathrm{r}\boldsymbol{y}_{\mathrm{up}} = 
\boldsymbol{P}_\mathrm{r}\boldsymbol{H}_{\mathrm{up}}\boldsymbol{x}_{\mathrm{up}} + \boldsymbol{P}_\mathrm{r}\boldsymbol{H}_{\mathrm{self}}\boldsymbol{P}_\mathrm{t}\boldsymbol{x}_{\mathrm{down}} + \boldsymbol{P}_\mathrm{r}\boldsymbol{n}_\mathrm{up},
}
\end{equation}
If we select the eigen-beamforming, by using $\boldsymbol{H}_{\mathrm{self}} = \boldsymbol{U}\boldsymbol{\Sigma}\boldsymbol{V}^\mathrm{H}$, we can obtain $\boldsymbol{P}_\mathrm{r} = \boldsymbol{S}_\mathrm{r}^\mathrm{T}\boldsymbol{U}^\mathrm{H}$ and  $\boldsymbol{P}_\mathrm{t} = \boldsymbol{V}\boldsymbol{S}_\mathrm{t}$, where $\boldsymbol{S}_\mathrm{r}^\mathrm{T} \in \left\{ 0,1\right\}^{N_\mathrm{up}\times M_\mathrm{up}}$ and $\boldsymbol{S}_\mathrm{t} \in \left\{ 0,1\right\}^{M_\mathrm{down}\times N_\mathrm{down}}$, where $N_\mathrm{up}$ and $N_\mathrm{down}$ are the number
of effective antennas for uplink and downlink, respectively \cite{Everett2016}. By minimizing the self-interference term, one strategy is to choose \cite{Riihonen2011}:
\begin{equation}
\label{Schoose}
{
\boldsymbol{S}_\mathrm{r}^\mathrm{T} = 
\left[
\begin{array}{ccc}
\boldsymbol{0} & \boldsymbol{I}_{N_\mathrm{up}}\\
\end{array}
\right]
~\mathrm{and}~ 
\boldsymbol{S}_\mathrm{t} = 
\left[
\begin{array}{c}
     \boldsymbol{0}\\
     \boldsymbol{I}_{N_\mathrm{down}}
\end{array}
\right].
}
\end{equation}
Therefore, we can obtain the power of self-interference term by \cite{Riihonen2011}
\begin{equation}
\label{Pselfintereference}
{
P_\mathrm{I} = P_\mathrm{down}\left\|\boldsymbol{S}_\mathrm{r}^\mathrm{T}\boldsymbol{\Sigma}\boldsymbol{S}_\mathrm{t}  \right\|_F^2 = 
P_\mathrm{down} \overset{\min{\left\{M_\mathrm{up},M_\mathrm{down}\right\} }}{\underset{i = M_\mathrm{up}+M_\mathrm{down}-\left(N_\mathrm{up}+N_\mathrm{down}\right)+1}\sum} {\sigma_\mathrm{self}^2[i]},
}
\end{equation}
where $\boldsymbol{R}_\mathrm{down} = \varepsilon \left\{\boldsymbol{x}_{\mathrm{down}}\boldsymbol{x}_{\mathrm{down}}^H\right\} = P_\mathrm{down}\boldsymbol{I}$; $\sigma_\mathrm{self}[i]$ is the singular value; $\varepsilon$ is the expectation operator.
And then, the desired signal term is 

\begin{equation}
\label{Psignal}
{
P_\mathrm{S} = \varepsilon \left\{\left\|\boldsymbol{P}_{\mathrm{r}}\boldsymbol{H}_{\mathrm{up}} \boldsymbol{x}_{\mathrm{up}}\right\|_2^2\right\}.
}
\end{equation}
If we assume that $\boldsymbol{R}_\mathrm{up} = \varepsilon \left\{\boldsymbol{x}_{\mathrm{up}}\boldsymbol{x}_{\mathrm{up}}^H\right\} = P_\mathrm{up}\boldsymbol{I}$, we can obtain $P_\mathrm{S} = P_\mathrm{up}\left\|\boldsymbol{P}_{\mathrm{r}}\boldsymbol{H}_{\mathrm{up}} \right\|_F^2$. 
The received noise level can be described by the dynamic range noise floor, the $i$-th antenna's noise level can be written as $P_{\mathrm{N},i} = \max{\left\{P_n,K(P_{\mathrm{S},i} + P_{\mathrm{I},i})\right\}}$ \cite{Everett2016}, where $P_n$ is the thermal noise floor based on channel bandwidth; $K$ is the noise dynamic range ratio; $P_{\mathrm{S},i} = P_\mathrm{up} \left|\left\{\boldsymbol{H}_{\mathrm{up}}\right\}_i\right|^2$;  $P_{\mathrm{I},i} = P_\mathrm{down} \left|\left\{\boldsymbol{H}_{\mathrm{self}}\boldsymbol{V}\boldsymbol{S}_\mathrm{t}\right\}_i\right|^2$. Here, we only consider $\boldsymbol{H}_{\mathrm{self}}$ is from the mutual couplings among antennas. The NLOS components because of the ambient bounces are neglected in our calculations since they are too small compared with the mutual couplings among antennas.


Finally, the total received $\mathrm{SINR}$ is 
\begin{equation}
\label{SINR_One}
{
\mathrm{SINR} = \frac{P_\mathrm{S}}{P_\mathrm{I}+P_\mathrm{N}}
}
\end{equation}

Here, we include the noise influence. The simplified SINR can be represented as 
\begin{equation}
\label{SINR_up}
{
\mathrm{SINR}_\mathrm{up} = \frac{P_\mathrm{up}\left\|\boldsymbol{S}_\mathrm{r}^\mathrm{T}\boldsymbol{U}^\mathrm{H}\boldsymbol{H}_{\mathrm{up}} \right\|_F^2}{P_\mathrm{down}\left\|\boldsymbol{S}_\mathrm{r}^\mathrm{T}\boldsymbol{\Sigma}\boldsymbol{S}_\mathrm{t}  \right\|_F^2+ \left\|\boldsymbol{S}_\mathrm{r}^\mathrm{T}\boldsymbol{U}^\mathrm{H} \boldsymbol{P}_\mathrm{N}\right\|_F^2}
}
\end{equation}
Accordingly, the downlink SINR can be represented by
\begin{equation}
\label{SINR_down}
{
\mathrm{SINR}_\mathrm{down} = \frac{P_\mathrm{down}\left\|\boldsymbol{H}_{\mathrm{down}}\boldsymbol{V}\boldsymbol{S}_\mathrm{t} \right\|_F^2}{\left\|\boldsymbol{P}_\mathrm{N,down}\right\|_F^2}
}
\end{equation}
where $\boldsymbol{P}_\mathrm{N,down}$ can be obtained by the same way as $\boldsymbol{P}_\mathrm{N}$ for uplink.

The reference SINR without SI reduction can be written as
\begin{equation}
\label{SINR_referenceup}
{
\mathrm{SINR}_\mathrm{up,ref} = \frac{P_\mathrm{up}\left\|\boldsymbol{H}_{\mathrm{up}}\right\|_F^2}{P_\mathrm{down}\left\|\boldsymbol{H}_{\mathrm{self}} \right\|_F^2 + \left\|\boldsymbol{P}_\mathrm{N}\right\|_F^2}
}
\end{equation}

\begin{equation}
\label{SINR_referencedown}
{
\mathrm{SINR}_\mathrm{down,ref} = \frac{P_\mathrm{down}\left\|\boldsymbol{H}_{\mathrm{up}}\right\|_F^2}{ \left\|\boldsymbol{P}_\mathrm{N,down}\right\|_F^2}
}
\end{equation}

For the in-band full-duplex, we consider also $M_\mathrm{up}=M_\mathrm{down}$ and there is no self-interference signal. The $i$-th antenna's noise level can be written as $P_{\mathrm{N,full},i} = \max{\left\{P_n,K\cdot P_{\mathrm{S},i}\right\}}$ 
\begin{equation}
\label{SINR_full-up}
{
\mathrm{SINR}_\mathrm{up,full} = \frac{P_\mathrm{up}\left\|\boldsymbol{H}_{\mathrm{up}}\right\|_F^2}{\left\|\boldsymbol{P}_\mathrm{N,full}\right\|_F^2}
}
\end{equation}

\begin{equation}
\label{SINR_full-down}
{
\mathrm{SINR}_\mathrm{up,full} = \frac{P_\mathrm{down}\left\|\boldsymbol{H}_{\mathrm{down}}\right\|_F^2}{\left\|\boldsymbol{P}_\mathrm{N,down,full}\right\|_F^2}
}
\end{equation}

When it comes to the half-duplex, we consider half of the time for uplink and the other half for downlink. 
\begin{equation}
\label{SINR_half_up}
{
\mathrm{SINR}_\mathrm{up,half} = \frac{P_\mathrm{up}\left\|\boldsymbol{H}_{\mathrm{up,half}}\right\|_F^2}{\left\|\boldsymbol{P}_\mathrm{N,half}\right\|_F^2}
}
\end{equation}
where the $i$-th antenna's noise level can be written as $P_{\mathrm{N,half},i} = \max{\left\{P_n,K\cdot P_{\mathrm{S},i}\right\}}$; $\boldsymbol{H}_{\mathrm{up,half}} \in \mathbb C^{\rm M}$;

\begin{equation}
\label{SINR_half_down}
{
\mathrm{SINR}_\mathrm{down,half} = \frac{P_\mathrm{up}\left\|\boldsymbol{H}_{\mathrm{down,half}}\right\|_F^2}{\left\|\boldsymbol{P}_\mathrm{N,down,half}\right\|_F^2}
}
\end{equation}
where $\boldsymbol{H}_{\mathrm{down,half}} \in \mathbb C^{\rm M}$;

When calculating the channel capacity, neglecting bandwidth, we can use
\begin{equation}
\label{capacity}
{
\mathrm{C}_\mathrm{down\& up} =\log_2({1+\mathrm{SINR}})
}
\end{equation}
to compute. It is noticed that we consider $\mathrm{C}_\mathrm{half} =\frac{1}{2}\log_2({1+\mathrm{SINR_\mathrm{half}}})$

From my understanding and tries, I think when the $M_\mathrm{up}=M_\mathrm{down}$, we can find the best solution as $N_\mathrm{up}< \frac{1}{2} M_\mathrm{down}$.  Within a limited area, $0.5\lambda$ spacing between adjacent elements is the best solution for this digital beamforming.

\bibliographystyle{IEEEtran}
\bibliography{ref.bib}

\end{document}